\documentclass[traditabstract]{aa}

\usepackage{url}
\usepackage[breaklinks=true]{hyperref}
\usepackage{twoopt}
\usepackage[english]{babel}          

\usepackage{natbib}
\bibpunct{(}{)}{;}{a}{}{,} 

\usepackage[utf8]{inputenc}
\usepackage[normalem]{ulem}
\usepackage{siunitx}
\usepackage{amsmath, amssymb}

\usepackage{multirow,bigstrut,ctable}
\usepackage[normalem]{ulem}
\usepackage{rotating}
\usepackage[varg]{txfonts} 
\usepackage{threeparttable}

\def \kms         {km$\,$s$^{-1}$}

\def \deg         {\text{$^{\circ}$}}
\def \arcmin      {\text{$^\prime$}}
\def \arcsec      {\text{$^{\prime\prime}$}}

\def \mjybeam     {mJy\,beam$^{-1}$}
\def \mujybeam    {$\mathrm{\mu}$Jy\,beam$^{-1}$}

\newcommand{\Hii}{\text{H\,\textsc{ii}}}
\newcommand{\Hi}{\text{H\,\textsc{i}}}
\newcommand{\Halpha}{\text{H\,$\alpha$}}

\newcommand{\beam}[2]{{#1}\arcsec$\times${#2}\arcsec}

\newcommand{\Msun}{\text{$\rm M_\odot$}}

\addto\extrasenglish{
    
}
\addto\extrasenglish{
    
}

\begin{document}

\title{The ViCTORIA project}
\subtitle{Description of a multi-frequency radio survey of the Virgo galaxy cluster}
\titlerunning{...}

\author{F.~de~Gasperin\inst{\ref{ira}}
\and H.~W.~Edler\inst{\ref{ham}}
\and A.~Boselli\inst{\ref{LAM}, \ref{cag}}
\and P.~Serra\inst{\ref{cag}}
\and M.~Fossati\inst{\ref{bic}}
\and V.~Heesen\inst{\ref{ham}}
\and A.~Merloni\inst{\ref{mpe}}
\and M.~Murgia\inst{\ref{cag}}
\and T.~H.~Reiprich\inst{\ref{bonn}}
\and A.~Spasic\inst{\ref{ham}}
\and N.~Zabel\inst{\ref{kap}}}
\authorrunning{F.~de~Gasperin et al.}

\institute{
INAF - Istituto di Radioastronomia, via P. Gobetti 101, 40129 Bologna, Italy \email{fdg@ira.inaf.it} \label{ira}
\and
Hamburger Sternwarte, Universit\"at Hamburg, Gojenbergsweg 112, 21029 Hamburg, Germany \label{ham}
\and
Aix Marseille Univ, CNRS, CNES, LAM, Marseille, France \label{LAM}
\and
Osservatorio Astronomico di Cagliari, Via della Scienza 5, 09047 Selargius (CA), Italy\label{cag}
\and
Dipartimento di Fisica G.Occhialini, Universit\`a degli Studi di Milano-Bicocca, Piazza della Scienza 3, 20126 Milano, Italy \label{bic}
\and
Max-Planck-Institut f\"ur extraterrestrische Physik (MPE), Giessenbachstrasse1, 85748 Garching bei M\"unchen, Germany\label{mpe}
\and
Argelander-Institut f\"ur Astronomie (AIfA), Universit\"at Bonn, Auf dem H\"ugel 71, 53121 Bonn, Germany\label{bonn}
\and
Kapteyn Astronomical Institute, University of Groningen, P.O.Box 800, 9700AV Groningen, The Netherlands\label{kap}
}

\date{Received ... / Accepted ...}

\abstract
{The Virgo cluster is the closest and richest nearby galaxy cluster. It is still in the formation process, with a number of sub-clusters undergoing merging and interactions. Although a great laboratory to study galaxy evolution and cluster formation, its large apparent size and the severe dynamic range limitations due to the presence of the bright radio source Virgo A (M\,87) reduced the ability of past wide-area radio surveys to image the region with high levels of sensitivity and fidelity.
In this paper we describe the Virgo Cluster multi-Telescope Observations in Radio of Interacting galaxies and AGN (ViCTORIA) project. The survey and its data reduction strategy are designed to mitigate the challenges of this field and deliver images from 42 MHz to 1.7 GHz frequencies of the Virgo cluster about 60 times deeper than existing data; final deliberables will include polarisation images and a blind \Hi{} survey aimed at mapping seven times more galaxies than previous experiments without selection biases.
Data have been collected with the Low-Frequency Array (LOFAR) using both the Low Band Antenna (LBA) and the High Band Antenna (HBA) systems and with MeerKAT in L-band, including polarisation and enough frequency resolution to conduct local \Hi{} studies.
At the distance of Virgo, current radio instruments have the resolution to probe scales of $\sim 500$~pc and the sensitivity to study dwarf galaxies, which are the most fragile systems given their shallow gravitational potential wells, making Virgo a unique laboratory to study galaxy evolution and AGN feedback in a rich environment. In this work, we present some preliminary results including the highest resolution images of the radio emission surrounding M\,87 ever captured that show that the lobes are filled with filamentary structures.
The combination of the presented radio surveys with state-of-the-art optical (NGVS, VESTIGE), UV (GUViCS), and X-ray (eROSITA) surveys will massively increase the scientific output from the studies of the Virgo cluster, making the ViCTORIA Project's legacy value outstanding.}


\maketitle

\section{Introduction}
\label{sec:introduction}

Being the closest rich cluster of galaxies \citep[16.5 Mpc;][]{Mei2007, Cantiello2018}, Virgo is an excellent laboratory to study environment-driven galaxy evolution, AGN feedback, galaxy-cluster formation, and dynamics. At the centre of the cluster lies the bright radio galaxy M\,87 \citep[Virgo \,A, NGC\,4486; e.g.,][]{Owen2000, deGasperin2012}. Virgo is closer than any other similarly massive cluster such as Hydra, Centaurus, and Abell 3558. Conversely, compared to other galaxy clusters at similar distance such as Fornax, Virgo has a halo mass about ten times larger, with estimates of the virial mass in the range of $M_{\rm vir} \approx 1.05 - 4.2 \times 10^{14}M_\odot$ \citep{McLaughlin1999, Urban2011, Simionescu2017, Boselli2022, McCall2024}. Virgo is the dominant concentration of mass in the local Universe and by convention is at the centre of the Local Supercluster \citep{Klypin2003}. The Virgo cluster is still in the formation process and dynamically young, with substructures visible both in the X-rays and in the galaxy distribution, such as interacting groups associated with the massive ellipticals M\,49, M\,60, and M\,86 \citep{Bohringer1994, Boselli2014a}. This is further confirmed by the high fraction of spiral galaxies \citep{Boselli2014a}, the significant deviation from spherical symmetry \citep{Binggeli1987}, and the properties of the intracluster stars \citep{Aguerri2005, Arnaboldi2005}.

In systems such as the Virgo cluster, the environment plays a major role in driving galaxy evolution. In fact, rich clusters of galaxies are predominantly populated by quiescent early-type galaxies, with a few star forming systems generally characterised by a low gas content and a reduced star formation activity \citep[e.g.][]{Dressler1980, BG2006}. Various mechanisms have been proposed to explain this different evolution \citep[e.g.][]{BG2006} from gravitational perturbations with other cluster members \cite[harassment;][]{Moore1998} to the hydrodynamic interaction of galaxies with the intracluster medium (ICM; ram pressure \citealt{Gunn1972, Boselli2022}; starvation \citealt{Larson1980}).

Furthermore, in such systems, the brightest cluster galaxies (BCGs), as well as other cluster members, may host powerful nuclear sources able to transfer mechanical energy to the surrounding ICM and modify its properties such as density and temperature, thus affecting the evolution of the entire cluster and its member galaxies \citep{Fabian2012}. Although all these mechanisms have been widely studied over the last 20 years, their duty cycle and the effects on galaxy evolution are still hotly debated. The main reason is that they involve complex physical processes acting on different scales, from giant molecular clouds and \Hii{} regions where star formation takes place ($\sim$50 pc) to the whole cluster ($\sim$1 Mpc). They also involve all gas phases, from the cold atomic and molecular hydrogen, to the ionised and hot gas, implying the need for high-quality multi-wavelength data covering a variety of environments.

\begin{figure}
        \centering
        \includegraphics[width=.5\textwidth]{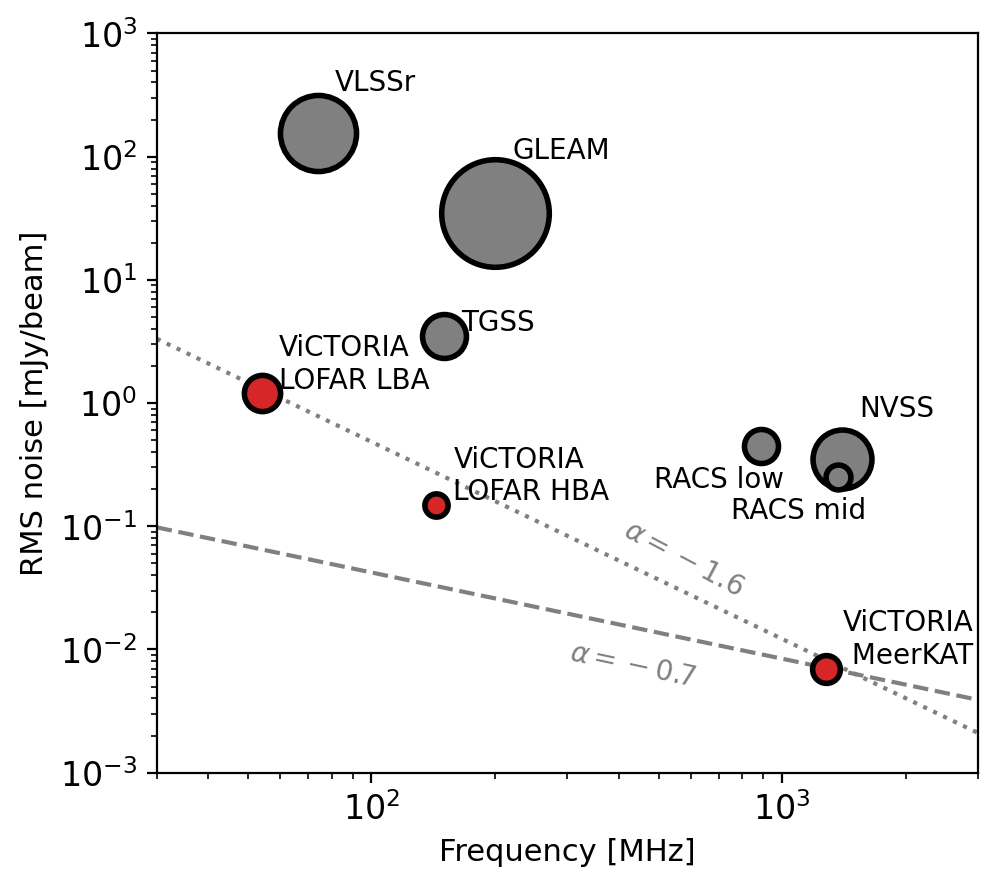}
        \caption{Frequency -- sensitivity plot comparing large radio surveys covering the Virgo cluster. The size of the markers scales linearly with resolution from 7\arcsec{} of ViCTORIA LOFAR HBA to 150 \arcsec{} of GLEAM. The sensitivity is derived from the Virgo region. References: GLEAM \citep[GaLactic and Extragalactic All-sky Murchison Widefield Array survey;][]{Hurley-Walker2017}, TGSS ADR1 \citep[TIFR GMRT Sky Survey - Alternative Data Release 1;][]{Intema2017}; VLSSr \citep[VLA Low-frequency Sky Survey redux;][]{Lane2014}; NVSS \citep[1.4 GHz NRAO VLA Sky Survey;][]{Condon1998}; RACS (Rapid ASKAP Continuum Survey) -low \citep{McConnell2020} and -mid \citep{Duchesne2023}.}
        \label{fig:sensitivity}
\end{figure}

Although it is an obvious target for large-scale radio surveys, the Virgo cluster has never been mapped with high-resolution, high-sensitivity blind radio surveys. The reason is twofold: firstly, the cluster angular size is large, with $r_{200}$ being about 5\deg{}.38 \citep[1.55\,Mpc;][]{Boselli2023b}\footnote{For consistency with NGVS and VESTIGE we assume throughout this paper $r_{200} = 1.55$~Mpc determined using the galaxy distribution; this value is larger than those presented in \citep{Boselli2022} and \citep{Simionescu2017} that are both close to $r_{200} = 1$~Mpc and have been derived using X-rays data.}, making surveys demanding in terms of observing time, especially at radio frequencies above 1 GHz; secondly, at the centre of the cluster lies M\,87, one of the brightest radio sources on the sky, whose radio emission reaches a flux density of 274~Jy at 1~GHz and~1700 Jy at~100 MHz, lowering the sensitivity of nearby regions because of dynamic range limitations. The most sensitive wide-area surveys that also cover the Virgo field, the TGSS ADR1 \citep[TIFR GMRT Sky Survey - Alternative Data Release 1;][]{Intema2017}, the NVSS \citep[1.4 GHz NRAO VLA Sky Survey;][]{Condon1998}, and the Rapid ASKAP Continuum Survey (RACS) -low \citep{McConnell2020} and -mid \citep{Duchesne2023} are all subject to severe imaging and calibration artefacts in the vicinity of M\,87.
Thanks to the use of two novel radio interferometers, the Low Frequency Array \citep[LOFAR;][]{VanHaarlem2013} and MeerKAT \citep{Jonas2016}, combined with advanced calibration and imaging techniques, we can complete a large-scale blind radio survey of the Virgo cluster with strongly improved image fidelity. With the Virgo Cluster multi-Telescope Observations in Radio of Interacting galaxies and AGN (ViCTORIA) project, we will deliver three blind radio surveys of the Virgo cluster spanning the frequency range from 42 to 1712 MHz and covering out to $r_{200}$. These radio images of the Virgo cluster region will be 60 times deeper than existing data (see Fig.~\ref{fig:sensitivity}), in full polarisation, and they will include a blind \Hi{} survey aimed at mapping seven times more galaxies than previous experiments and without selection biases. The combination with state-of-the-art optical \citep[NGVS, VESTIGE;][]{Ferrarese2012, Boselli2018a}, UV \citep[GUViCS;][]{Boselli2011}, and X-ray \citep[eROSITA;][]{McCall2024} surveys will massively increase the scientific potential of this dataset, making its legacy value outstanding.

Throughout this paper, we assume a flat $\mathrm{{\Lambda}CDM}$ cosmology with $\Omega_\mathrm{m}=0.3$ and $H_0=70\,\mathrm{km\,s^{-1}\,Mpc^{-1}}$. At the distance of M\,87, for which we adopt a value of $16.5\pm0.1$ (random) $\pm1.1$ (systematic)\,Mpc (\citealt{Mei2007}; see also \citealt{Cantiello2018, Cantiello2024}), one arcsecond corresponds to 80\,pc. This paper is arranged as follows. In \autoref{sec:observations}, we present the observations that were or will be released under the project's umbrella, as well as an overview of the other blind surveys that covered the Virgo cluster at complementary wavelengths. In \autoref{sec:comparisons}, we present comparisons with current surveys as well as the highest resolution images of Virgo A (M\,87) to date at the frequencies of 54, 144, and 1284 MHz. In \autoref{sec:science} we outline the science cases that are driving the project, and we conclude in \autoref{sec:conclusions}.

\section{Strategy and observations}
\label{sec:observations}

\begin{table*}
\caption{Main properties of the set of surveys that are part of the ViCTORIA Project.}\label{tab:surveys}
\centering
\begin{threeparttable}
\begin{tabular}{lccccccc}
\hline
Telescope & Frequency range & Resolution & Sensitivity & \# pointings & Area\tnote{a} & Tot. int. time\tnote{b} & Projects\\
          & (MHz)           & (arcsec)   & (\mujybeam{}) & & (deg$^2$) & (hrs) & \\
\hline
LOFAR LBA & 42 -- 66 & 17 & 1700 & 9 & 166 & 24\tnote{c} + 36\tnote{c} & LC18\_012, LC20\_038 \bigstrut[t]\\
LOFAR HBA & 120 -- 168 & 7 & 140 & 9 & 132 & 64\tnote{d} & LC11\_010\\
MeerKAT & 856 -- 1712 & 10 & 7 & 140 + 180 & 112 & 100 + 130 & SCI-20230907-FD-01,\\
&&&&&&& SCI-20220822-FD-02\\
\hline
\end{tabular}
\begin{tablenotes}
    \item[a] Measured considering a primary beam response ${\geq}30\%$.
    \item[b] On-target time.
    \item[c] Quadruple beam with a beam on the calibrator 3c295; each field was observed for 18 hrs, and the field of M87 was observed for 36 hrs. 
    \item[d] Dual beam with a beam on M87; each Virgo field was observed for 8 hrs, and the field of M87 was observed for 64 hrs.
\end{tablenotes}
\end{threeparttable}

\end{table*}

\begin{figure}
        \centering        
        \includegraphics[width=.5\textwidth]{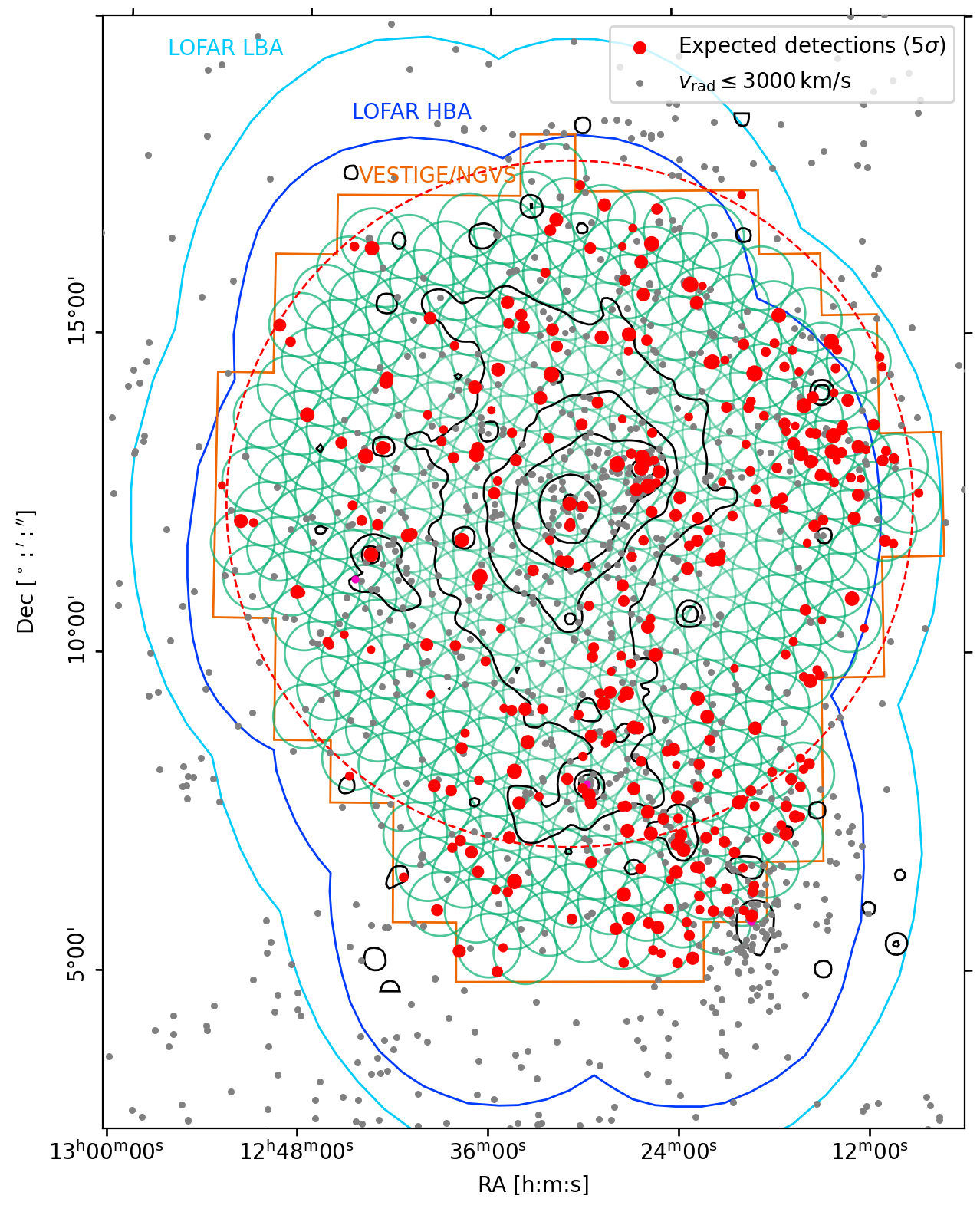}
        \caption{Coverage of ViCTORIA project surveys. In green are the pointings at mid-frequency FWHM of the MeerKAT Virgo Survey designed to fill the orange region covered by NGVS in optical and VESTIGE in \Halpha. In light and dark blue are the coverages of LOFAR LBA and HBA, respectively (mid-frequency FWHM). The red dashed line is the estimated $r_{200}$. The contours trace the X-ray emission from ROSAT \citep{Bohringer1994}, grey dots are VCC galaxies with recessional velocity $\leq 3000$~\kms{}, and red dots mark cluster galaxies that we expect to detect based on their \Halpha-brightness.}
        \label{fig:coverage}
\end{figure}

The ViCTORIA project consists of three radio surveys that span the 42 -- 1712 MHz frequency range. The surveys are carried out using the LOFAR Low Band Antenna (LBA) and High Band Antenna (HBA) systems, and MeerKAT data were taken with the L-band receivers. Basic data on the surveys are summarised in Table~\ref{tab:surveys}. The coverage of the three surveys is visible in Fig.~\ref{fig:coverage}, and the joint area is about 99 deg$^2$.

\subsection{ViCTORIA radio surveys}
\label{sec:surveys}

Here, we describe the observing strategies and preliminary results of the ViCTORIA surveys in more detail.

\subsubsection{The LOFAR LBA Virgo Cluster Survey}

The survey covers 166~deg$^2$ and encompasses the entire Virgo cluster to more than $r_{200}$. The survey footprint (see Fig.~\ref{fig:coverage}) is composed of nine telescope pointings, with one central pointing aligned with M\,87 and eight others covering the outer regions of the cluster. The survey area is elongated towards the south, following the distribution of the ICM and cluster galaxies. The data were collected in two rounds of observations, the first during 2022 and the second in 2023, which finished in early 2024. The total observation time is 60\,h, and observations are taken exploiting LOFAR's multi-beam capability with four parallel beams in the $42-66$\,MHz frequency range. One beam is always pointed on the calibrator source 3C\,295, and the other three observe the selected Virgo Cluster fields. Fields that are observed simultaneously were chosen to minimize the overlap and, hence, the amount of correlated noise. The resulting effective exposure is 36\,hrs for the M\,87 field and 18\,hrs for each of the flanking fields.

The presence of the extremely bright ($S_\mathrm{54MHz}=2.6$\,kJy), extended (15\arcmin), and morphologically complex radio source M\,87 presents a challenge for the data reduction of these observations. Further severe difficulties arise due to the low declination of the field (${\sim}10\deg$), which increases the field of view (FoV) and the projected ionospheric electron density since observations are taken at rather low elevations of $25\deg{-}50\deg$. Both of these effects increase the direction-dependent variation of the ionospheric corruptions across the FoV \citep[see][]{deGasperin2018a}.

For M\,87 itself, a high-fidelity image at 54\,MHz was obtained previously \citep{deGasperin2020}. For the calibration of interferometric LOFAR LBA observations, the standard calibration procedures are available in the Library for Low-Frequencies (LiLF\footnote{https://github.com/revoltek/LiLF}; \citealt{deGasperin2019, deGasperin2020A, deGasperin2021}). To address the additional complications of the Virgo field, a `peeling' strategy to subtract M\,87 from the visibility data prior to the direction-dependent calibration will be employed. We followed a similar procedure for the calibration of the LOFAR HBA Virgo Cluster Survey (see \citealt{Edler2023} and the following paragraph).

We simulated the expected quasi-thermal noise level of the full mosaic based on the source-equivalent flux densities provided in \citet{VanHaarlem2013}, taking into account the pointing overlap and sensitivity loss due to the declination and primary beam attenuation. The median quasi-thermal noise level across the full LBA survey area is $\approx2.2$~\mjybeam{}, and $\approx1.7$~\mjybeam{} for the sky area that is common among the LBA, HBA, and MeerKAT surveys. The resolution of the final mosaics will be approximately $25''\times15''$.



\subsubsection{The LOFAR HBA Virgo Cluster Survey}

The LOFAR HBA Virgo Cluster Survey \citep{Edler2023} covers a 132\,deg$^2$ area of the Virgo cluster. The pointing scheme follows the same nine fields as the LBA survey, except that each observation has a smaller FoV due to the different station layout and observing frequency. The covered region extends out to $r_{200}$. The total observation time of 64\,hrs was collected between March 18 2019 and April 2 2019. Two parallel target beams were used, one on M\,87 and the other cycling the rest of the fields, in the frequency range from $120-168$\,MHz. The observations were carried out in blocks of 8\,hrs each, yielding an effective exposure of 64\,hrs for the field containing M\,87 and 8\,hrs for the eight outer fields. At the beginning and end of each block, a calibrator source (either 3C\,196 or 3C\,295) was observed for 10 min. This observational setup is identical to the observations of the LOFAR Two-metre Sky Survey \citep[LoTSS;][]{Shimwell2019,Shimwell2022}. However, due to the presence of M\,87, an adjusted calibration strategy was necessary. After deriving the corrections for the instrumental systematic effects in the calibrator observation and applying them to the target data set, a specialized \emph{\textit{peeling}} step was carried out to take care of M\,87. In this step, the observations were phase-shifted towards M\,87 and averaged in time and frequency. Afterwards, we solved against a high-quality source model of M\,87, taking into account the direction-dependent variation of the primary beam across M\,87. In this step, we derived scalar phases on 16\,s time intervals and full-Jones matrices on 256\,s time intervals. These solutions were used to subtract M\,87 from the full-resolution data set, which was then phase-shifted back to the original phase centre. Subsequently, \texttt{ddf-pipeline} \citep{Tasse2021} was used for the full direction-dependent calibration of the observation. 

These steps were repeated for each target field, and the resulting images were mosaicked. The detailed description of the calibration, survey properties, and data products of the survey can be found in \citet{Edler2023}. In the final survey mosaic, 112 Virgo cluster member galaxies were detected at a significance level of ${>}4\sigma$. The catalogue and image data of the LOFAR HBA Virgo Cluster survey are available at the CDS\footnote{\url{https://cdsarc.cds.unistra.fr/viz-bin/cat/J/A+A/676/A24}} as well as on the LOFAR Surveys Key Science Project website\footnote{\url{https://lofar-surveys.org/virgo_data.html}}.

Mosaics were created at three different resolutions: $9''{\times}5''$ (high), $20''$ (low) and $1'$ (very low). At high resolution, which is slightly lower than the one of LoTSS due to the low declination of the Virgo cluster, the median noise level is $\sigma_\mathrm{rms}=140\,\mathrm{\mu{Jy}\,beam^{-1}}$, while for the full area, $\sigma_\mathrm{rms}=280\,\mathrm{\mu{Jy}\,beam^{-1}}$ is reached.

\subsubsection{The MeerKAT Virgo Cluster Survey}

The MeerKAT Virgo Cluster Survey covers an area of 112\,deg$^2$, extending to about $r_{200}$ and following the footprint of the optical surveys VESTIGE and NGVS (see \autoref{sec:complementary}). The survey was completed in two different cycles, the first observed in 2023 covering the inner region up to about 1 Mpc in radius, and the second observed in 2024 expanding to the VESTIGE/NGVS footprints. The total observing time is 230~hrs on target and 287~hrs including overheads. Each observing run included observations on the bandpass calibrator (J1939-6342) and the polarisation calibrator (3C\,286). The target observations were interleaved with ten-minute-long observations of a gain calibrator (J1150-0023). The total number of pointings is 320, and they are organised in a hexagonal grid with a spacing of 0.58\deg{} yielding a full Nyquist-sampling ($\Theta_{\mathrm{PB}}/\sqrt{3}$) at the \Hi{} frequency (1420 MHz) and closer than $\Theta_{\mathrm{PB}}/\sqrt{2}$ at the lowest end of the band. Each pointing has a total of 43 -- 45 minutes of effective integration time spanning a 4.5 hr observing run.

\begin{figure}
        \centering 
        \includegraphics[width=.45\textwidth]{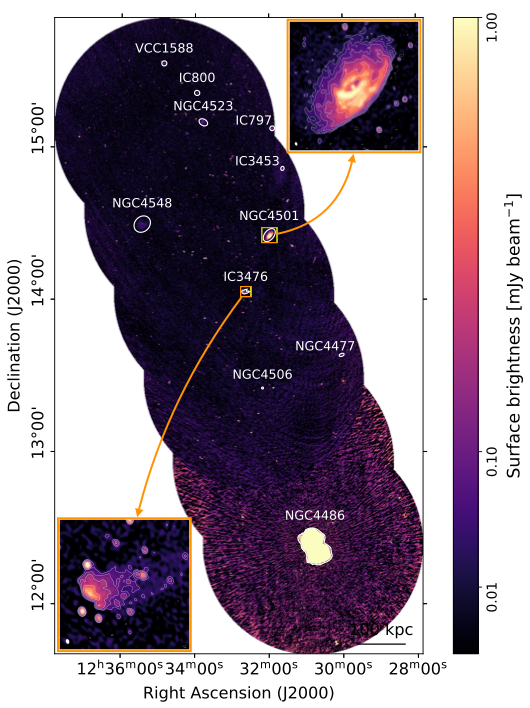}
        \caption{MeerKAT L-band data mosaic of five pointings covering M87 and moving towards the north-east. These fields were used as a pilot experiment to derive a good calibration strategy. The dynamic range limitation due to the presence of the bright radio emission from M87 is evident in the increased rms noise in the pointings where the source is within the primary beam.}
        \label{fig:mosaic}
\end{figure}

Dynamic range limitation due to the presence of the bright source M\,87 has been proven to be under control thanks to a pilot experiment that covered five fields moving away from the bright source (see Fig.~\ref{fig:mosaic}). The observing strategy of the pilot pointings is equal to the one used for the rest of the survey. Thanks to the pilot experiment, we were able to derive an ad hoc calibration strategy to properly remove (peel) the emission of M\,87 from the data for the affected fields. With the pilot we showed that only the closer pointings ($<1\deg$) have an increased noise level.

The expected noise and resolution, confirmed by the pilot experiment, of the survey is 7~\mujybeam{} at a resolution of 7.6\arcsec{} (Briggs: $-0.5$). The line noise is 0.204~\mujybeam{} per 25~\kms{} channel at a 30\arcsec{} resolution.

\subsection{Main deliverables}

Using the aforementioned data, the ViCTORIA project is aimed at producing the following deliverables:

   \begin{table*}
   {\centering
      \caption[]{ViCTORIA \Hi\ mosaic cubes}
         \label{table:mos}
         \begin{tabular}{llllllll}
            \noalign{\smallskip}
            \hline
            \hline
            \noalign{\smallskip}
             label & restoring beam$^\dagger$ & channel width & Briggs \emph{robust} & $uv$ taper & pixel size & noise$^{\dagger \dagger}$ & $N(\Hi)_{3\sigma, 25\mathrm{km/s}}$ \\
              & ($B_\mathrm{maj}\times B_\mathrm{min}$, $B_\mathrm{PA}$) & (\kms) &  &  &  & (mJy beam$^{-1}$) & (cm$^{-2}$) \\
            \noalign{\smallskip}
            \hline
            \noalign{\smallskip}
            $10''$  & $12.0''\times8.0''$,   190$^\circ$ & 5.5 & 0.0    & $0''$   & $2''$   & 0.64 & $2.6\times10^{20}$ \\
            $15''$  & $17.2''\times13.7''$,  165$^\circ$ & 5.5 & 0.0    & $9''$   & $3''$   & 0.59 & $9.7\times10^{19}$ \\
            $27''$  & $30.5''\times24.5''$,  165$^\circ$ & 5.5 & 0.0    & $17''$  & $5''$   & 0.67 & $3.5\times10^{19}$ \\
            $47''$  & $52.0''\times43.0''$,  170$^\circ$ & 5.5 & 0.5    & $30''$  & $10''$  & 0.63 & $1.1\times10^{19}$ \\
            $79''$  & $84.0''\times 74.0''$,  20$^\circ$ & 5.5 & 0.5    & $64''$  & $20''$  & 0.77 & $4.8\times10^{18}$ \\
            $125''$ & $130.0''\times120.0''$, 20$^\circ$ & 5.5 & 1.0   & $112''$ & $36''$  & 1.03 & $2.6\times10^{18}$ \\
            \noalign{\smallskip}
            \hline
            \noalign{\smallskip}
         \end{tabular}

         \small \emph{Notes.} ($\dagger$) Each channel of each ViCTORIA field has its own $uv$ coverage and therefore its own dirty beam. As in \cite{Serra2023}, we adopted a fixed restoring beam for all channels and all fields to simplify the analysis of the \Hi\ mosaics. The parameters of these fixed restoring beams are shown in this table. They were chosen as the best match to the ensemble of all dirty beams of the first 135 ViCTORIA fields observed. ($\dagger \dagger$) Median noise level per channel across the mosaic. The actual noise varies as a function of position and frequency.\normalsize
         }
   \end{table*}

\begin{itemize}
    \item Three \textbf{continuum surveys} at 42 -- 66 MHz, 120 -- 168 MHz, and 856 -- 1712 MHz, reaching the sensitivity levels of 1700, 150, and 7~\mujybeam{}, respectively. No other large survey of the Virgo cluster has been attempted at these frequencies, so the improvement compared to NVSS at 1.4 GHz will be of a factor of 60 in depth and six in resolution (see Fig.~\ref{fig:sensitivity}). Based on VESTIGE and the \Halpha{}-radio correlation \citep{Boselli2015}, the high-frequency survey should detect $\sim 330$ galaxies (vs. 79 known before) entering -- for the first time -- well into the parameter space of dwarf systems. The data will also enable the detection of low-surface-brightness features associated with the stripped material accessible only through a blind survey, and reach beyond $r_\mathrm{vir}$ of a $M_{200}\geq 10^{14}~\Msun$ cluster at this depth and spatial resolution for the first time. Specifically, the large coverage of the ViCTORIA surveys will enable the detection of those galaxies entering into the cluster for the first time; here, ram pressure is likely to be an efficient mechanism for their transformation \citep{Boselli2022, Boselli2023b}.

    \item A \textbf{blind \Hi{} survey} with an $N(\Hi)$ sensitivity level (3$\sigma$ over a line width of 25~\kms{}) from $\sim 2 \times 10^{18}$~cm$^{-2}$ at $\sim2'$ resolution ($\sim10$ kpc at the distance of Virgo) to $\sim2 \times 10^{20}$~cm$^{-2}$ at $\sim10''$ resolution ($\sim 0.8$ kpc; see Table \ref{table:mos}). This combination of resolutions and column densities matches the predictions of numerical hydrodynamical simulations for the density of the gas in the tails of stripped material \citep{Tonnesen2010} that are in close agreement with other observations \cite{Chung2007}. At a high resolution, similar quality data are only available for 53 massive galaxies \citep[VIVA;][]{Chung2009}. The full MeerKAT survey will bring the \Hi{} detections of resolved galaxies to $\sim 370$, extending the dynamic range to dwarf systems, reaching $M_\Hi = 2.0 \times 10^6~\Msun$ (3$\sigma$ over a 100~\kms line width).
    
    \item A \textbf{blind polarisation survey} aimed at detecting the magnetic field structure in galaxies, tails, and AGN-driven emission. The data will allow the identification of an estimated 2000 objects through RM synthesis (priv. comm. MIGHTEE collaboration).
    
    \item A \textbf{resolved spectral index} estimation between 144 MHz (LOFAR HBA) and 1284 MHz (MeerKAT) of all radio sources with $S_{\rm 144\,MHz} > 450$~\mujybeam{}, at an angular scale of 7\arcsec{} (560 pc), and down to spectral index $\alpha=-1.4$. Including the data at 54 MHz (LOFAR LBA), the study of spectral curvature will be possible at a resolution of 17\arcsec{} (1.4 kpc).
    
\end{itemize}

\subsection{Complementary surveys}
\label{sec:complementary}

The favourable declination of Virgo ($\sim 12\deg$), accessible from both hemispheres, enabled several high-profile multi-$\lambda$ surveys. The whole cluster was observed in the 1980s with the du Pont 2.5m telescope at Las Campanas by Sandage, Tammann, and Binggeli, who made the first systematic identification of the Virgo cluster members, the Virgo Cluster Catalogue \cite[VCC][]{Binggeli1985}, and several targeted studies of the statistical properties of its galaxies \citep[e.g.][]{Sandage1985}. Since this seminal work, which was still based on photographic plates, different blind surveys covered the whole cluster. In the optical band, after the SDSS observations of the cluster \citep[e.g.][]{Lisker2006a, Lisker2006b, Lisker2007} and of its surrounding regions leading to the Extended Virgo Cluster Catalogue \citep[EVCC;][]{Kim2015}, the cluster was the target of the Next Generation Virgo cluster Survey \citep[NGVS;][]{Ferrarese2012} undertaken with MegaCam at the Canada French Hawaii Telescope (CFHT). The very deep $u,g,i,z$ images, which covered 104 deg$^2$ corresponding to $\simeq r_{200}$, allowed the identification of $\simeq 3700$ members \citep{Ferrarese2020} and the study of their structural properties down to the dwarf galaxy population ($M_{\rm star}$ $\sim$ 10$^6$ M$_{\odot}$), including those of peculiar objects such as ultra-diffuse galaxies \citep[UDGs;][]{Lim2020, Junais2022} and ultra-compact galaxies \citep[UCDs;][]{Liu2020}. The cluster was observed up to $R \simeq 2 r_{200}$ in the FUV ($\lambda$ = 1539 \AA) and NUV ($\lambda$ = 2316 \AA) bands during the GALEX Ultraviolet Virgo Cluster Survey \citep[GUViCS;][]{Boselli2011}. At these wavelengths, the emission of galaxies is dominated by the young stellar population \citep[e.g.][]{Kennicutt1998, Boselli2009}. The cluster was also covered by the Virgo Environmental Survey Tracing Ionised Gas Emission \citep[VESTIGE;][]{Boselli2018}, a blind narrowband H$\alpha$ imaging survey carried out with MegaCam at the CFHT. This survey allowed the detection of all the star forming galaxies of the cluster (384 objects) and the estimation of their present day star formation activity \citep{Boselli2023a, Boselli2023b}. The GALEX and the VESTIGE data were thus fundamental for reconstructing the star formation history of all cluster members and finally studying the quenching process occurring in dense environments with unprecedented accuracy \citep[e.g.][]{Boselli2015, Boselli2023c}. The dust emission of the cluster galaxies was secured thanks to the all-sky WISE survey \citep{Wright2010} and the Herschel Virgo cluster Survey \citep[HeViCS;][]{Davies2010, Auld2013}, a blind survey covering the central 64 deg$^2$ of the cluster at 100, 160, 250, 350, and 500 $\mu$m. Dedicated X-ray observations such as those done with Rosat \citep{Bohringer1994}, XMM-Newton \citep{Urban2011}, Suzaku \citep{Simionescu2017}, and eROSITA \citep{McCall2024} were crucial for deriving the main properties of the intracluster medium (ICM; hot gas distribution, density, and temperature), and compare them to those of different cluster members to quantify the perturbations induced by the surrounding environment on galaxy evolution \citep{Vollmer2001, Gavazzi2013, Boselli2014}. In the radio domain, the cluster was covered by several wide area radio-continuum surveys (see Sec.~\ref{sec:radiosurveys}) and in the \Hi{} line at 21 cm by HIPASS \citep{Meyer2004} and by ALFALFA \citep{Giovanelli2005}.

Thanks to its proximity, many Virgo cluster galaxies were and are still the targets of dedicated observations at almost all wavelengths. Worth mentioning are the ACS Virgo Cluster Survey of 100 early-type galaxies with the Hubble telescope \citep{Cote2004}, VLA \Hi{} \citep[VIVA;][]{Chung2009}, Arecibo \Hi{} \citep{Gavazzi2005}, ALMA CO \citep[VERTICO;][]{Brown2021}, and VLT/MUSE IFU spectroscopy (MAUVE, VLT Large program, PI Cortese) observations of 50 late-type galaxies or pointed observations in X-ray with Einstein \citep{Fabbiano1992, Shapley2001} and Chandra \citep{Gallo2008, Soria2022}, and mid- and far-IR with ISO \citep{Leech1999, Tuffs2002, Boselli2003}, Spitzer \citep{Bendo2012}, and Herschel \citep{Ciesla2012, Cortese2014}. Pointed observations of individual galaxies are also available in the radio continuum at 4.8, 8.6, and 10.55 GHz \citep{Niklas1995, Vollmer2004, Capetti2009}.  All cluster members with a $B$-band magnitude brighter than $m_B \lesssim 16$ AB mag have been observed in near-infrared bands \citep{Boselli1997, McDonald2011, Munoz2014, Janz2014} and the star forming one with pointed \Halpha{} narrowband imaging \citep{Koopmann2001, Gavazzi2002, Gavazzi2013}.



\section{Comparisons with current surveys}
\label{sec:comparisons}

\subsection{Comparison with H\textsc{i} surveys}

\begin{figure}
        \centering 
        \includegraphics[width=.49\textwidth]{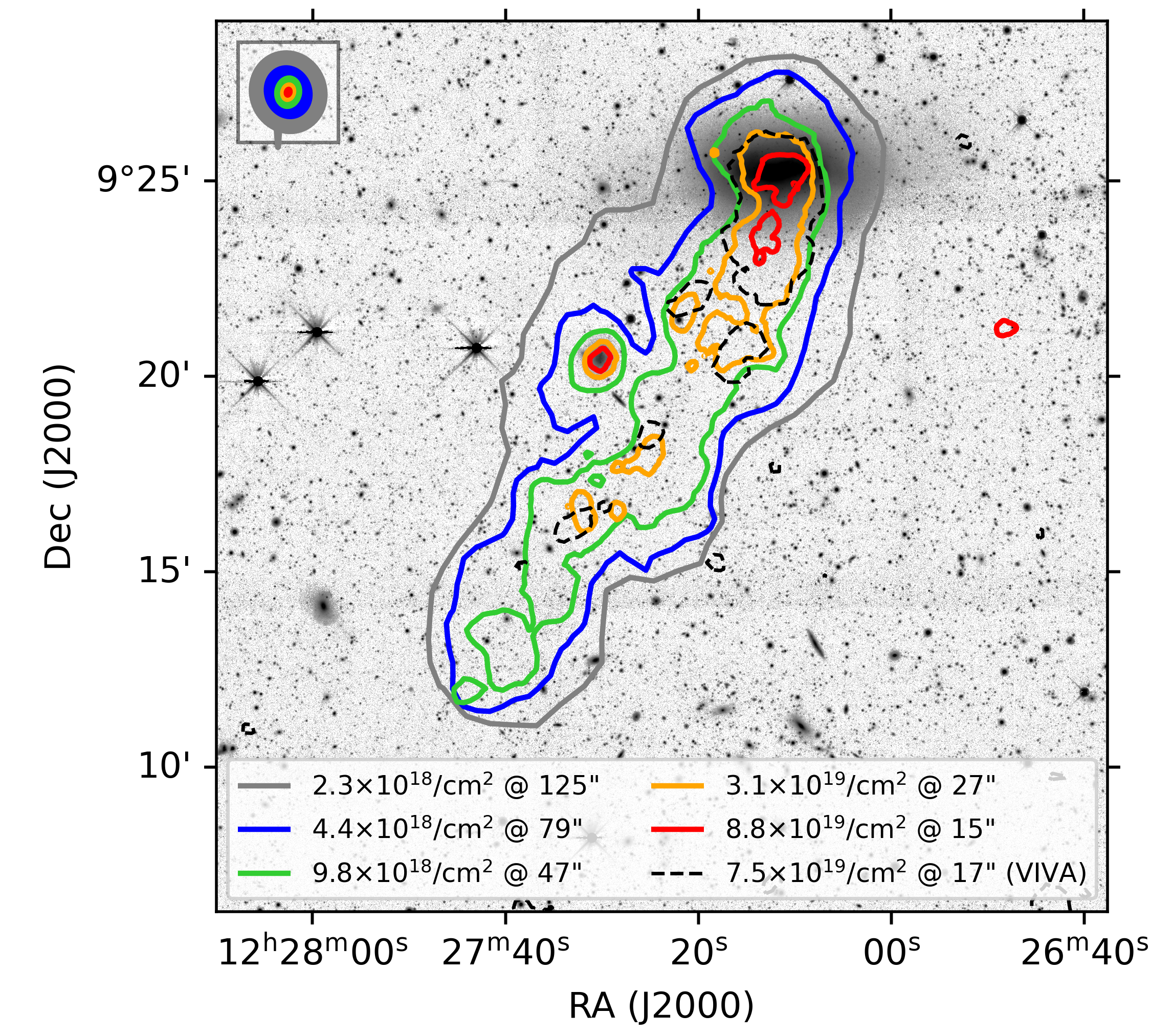}
        \caption{\Hi{} gas distribution of perturbed galaxy NGC~4424 seen by ViCTORIA. For each resolution, we show the lowest reliable contour (3$\sigma$ over a line width of 25 km s$^{-1}$). The contours are overlaid on an $r$-band image downloaded from the Legacy Survey database. The contour levels and resolutions are listed in the legend at the bottom, and the resolutions are shown in the top left corner. For comparison, we also show the lowest reliable \Hi\ contour from the VIVA survey \citep{Chung2009}, defined for consistency at a 3$\sigma$ level over a line width of 25 km s$^{-1}$.}
        \label{fig:NGC4424}
\end{figure}

The VIVA survey \citep{Chung2009} provided pointed observations of 53 galaxies in the Virgo cluster, most of which are located within the footprint of the ViCTORIA observations. The typical $3\sigma$ sensitivity level of the VIVA data is $N(\Hi) \sim 6 - 8 \times 10^{19}$~cm$^{-2}$ at an angular resolution of $\sim 15\arcsec$ and assuming a line width of 25~\kms{}. We can thus compare the two different sets of data. Fig.~\ref{fig:NGC4424} shows the galaxy NGC~4424, which is located in the southern portion of the Virgo cluster. This galaxy exhibits a perturbed \Hi{} morphology, with an extended low-column-density tail in the south-east direction, this is indicative of recent interactions. The figure shows that with ViCTORIA we recover all the emission detected by VIVA at a comparable resolution. However, we also detect significantly more diffuse \Hi thanks to the much better \Hi{} gas-column-density sensitivity reached at a lower resolution. NGC 4424 was also observed using the Karoo Array Telescope (KAT-7), a precursor of MeerKAT, with 30 hrs of on-source integration, and with the Westerbork Synthesis Radio Telescope (WSRT) using APERITIF (12 hrs on-source; \cite{Sorgho_2017}). The extended tail was detected in the combined dataset with properties of ($N(\Hi) \sim 5 \times 10^{18}$~cm$^{-2}$, 3$\sigma$ at an angular resolution of $\sim 174 \arcsec \times 30 \arcsec$, and assuming a line width of 16.5~\kms{}) similar to those gathered in the MeerKAT data (see Fig. 5 in \cite{Sorgho_2017}). ViCTORIA is thus perfectly suited to searching for extended, low \Hi{} gas column densities such as those produced during the interaction of galaxies with their surrounding environment, while still enabling the study of the \Hi\ morphology and kinematics at a high resolution. 

\begin{figure}
        \centering 
        \includegraphics[width=.49\textwidth]{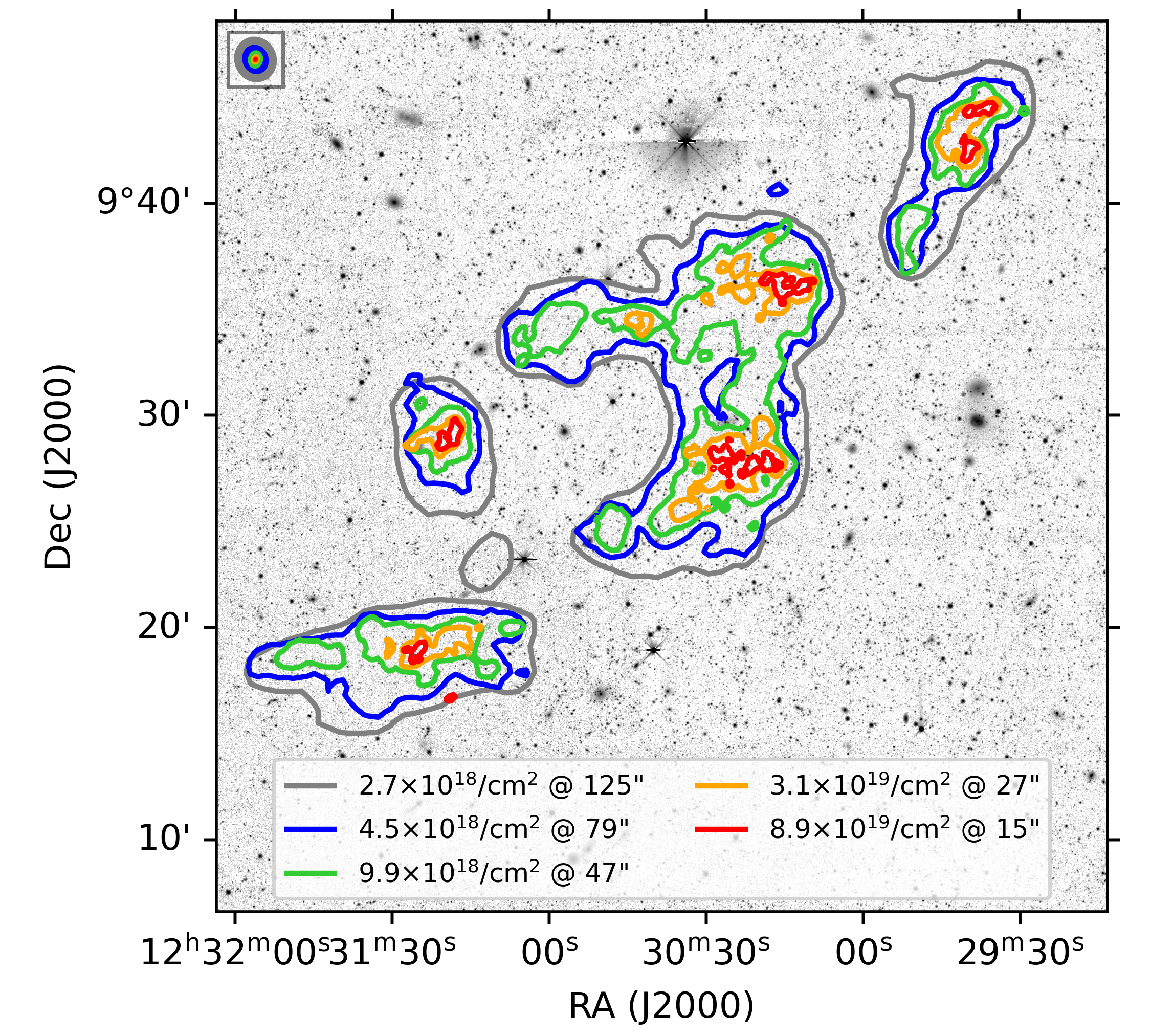}
        \caption{\Hi{} clouds detected during ALFALFA survey by \cite{Kent2009} as seen by ViCTORIA. For each resolution, we show the lowest reliable contour (3$\sigma$ over a line width of 25 km s$^{-1}$). The contours are overlaid on an $r$-band image downloaded from the Legacy Survey database. The contour levels and resolutions are listed in the legend at the bottom, and the resolutions are shown in the top left corner.}
        \label{fig:k09clouds}
\end{figure}

The ALFALFA survey \citep{Giovanelli2005} covered the whole Virgo cluster region with the Arecibo radio telescope, detecting, within the ViCTORIA footprint, 267 galaxies \citep{Haynes2018} and several \Hi{} clouds not associated with any stellar counterpart \citep{Kent2007,Kent2009}\footnote{A blue, compact star forming complex has been recently identified by \cite{Jones_2024} as the probable stellar counterpart of the cloud complex discovered in \citep{Kent2009}}. The typical ALFALFA sensitivity level is $N(\Hi) = 2 \times 10^{18}$ cm$^{-2}$ for \Hi{} filling the 3.5\arcmin{} beam and assuming a line width of 25~\kms{} with channels of 10~\kms{} \citep[see Sect. 6 of][]{Giovanelli2005}. This low \Hi{} column density can be reached by the ViCTORIA survey, as shown in Fig. \ref{fig:k09clouds} for the complex of clouds originally discovered by \cite{Kent2009} and later studied by \citep{Jones_2024}. All of the five clouds discovered by ALFALFA are detected in the MeerKAT data. Mapped at higher angular resolution ($\sim$ 125 \arcsec vs $\sim$ 210 \arcsec), they show a different morphology . The higher angular resolution of ViCTORIA compared to ALFALFA thus enables a more detailed analysis of their morphology and kinematics.

\subsection{Comparison with blind radio-continuum surveys}
\label{sec:radiosurveys}

\begin{figure*}
        \centering 
        \includegraphics[width=\textwidth]{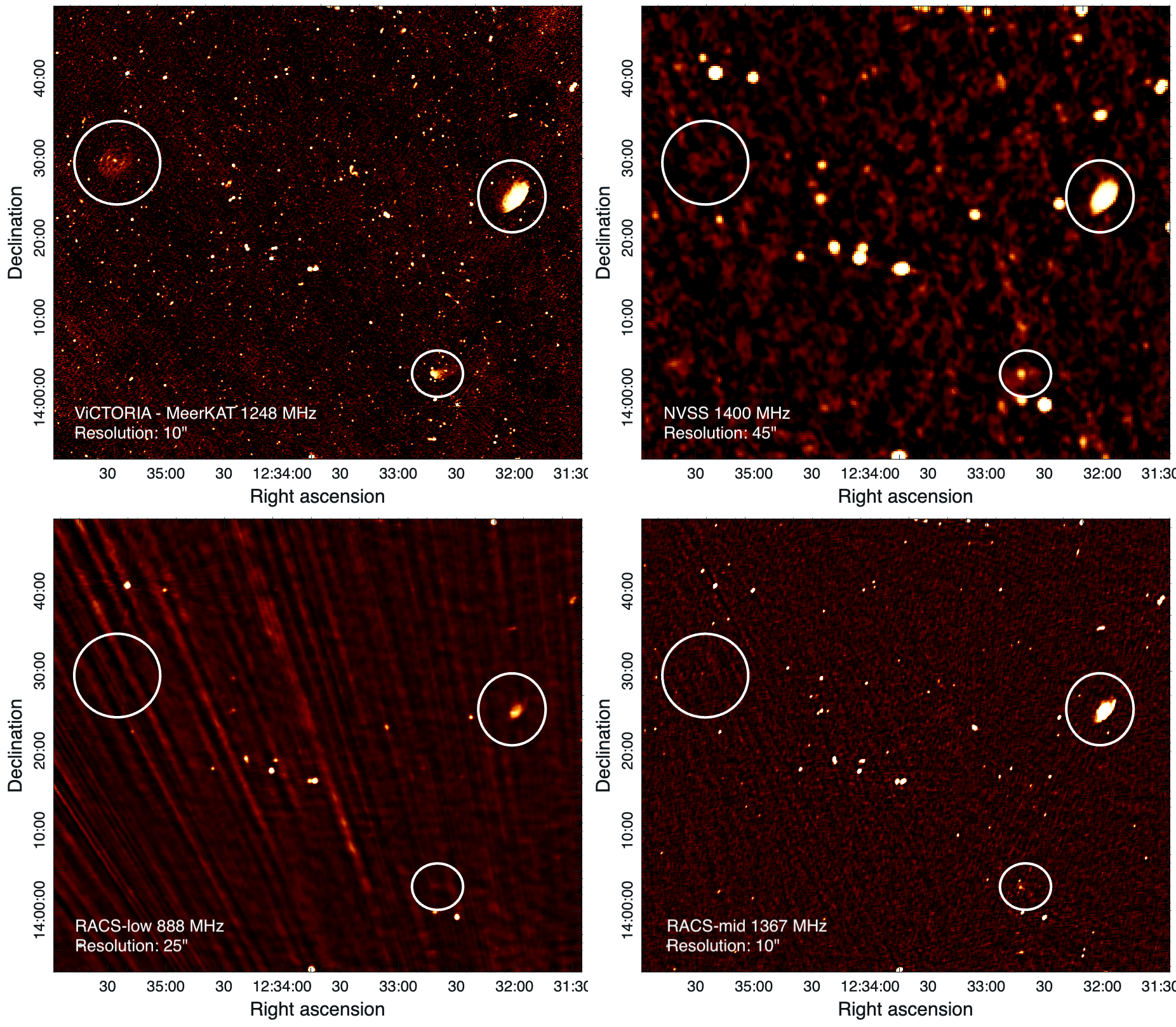}
        \caption{Comparison of various surveys in L band: ViCTORIA MeerKAT (from the pilot fields shown in Fig.~\ref{fig:mosaic}), NVSS, and RACS low and mid. The circles show the locations of NGC~4548, NGC~4501 and IC~3476 (from top left to bottom right).}
        \label{fig:contcomp}
\end{figure*}

Blind radio surveys of the Virgo cluster are complicated to perform mostly because of the wide apparent size of the cluster that, including the substructures, covers more than 100~deg$^2$ of the sky and because of the presence of the radio source Virgo~A (M87). The source is among the brightest in the northern sky, reaching a flux density of $\sim 200$~Jy at 1400~MHz and $\sim 1300$~Jy at 140~MHz. The presence of Virgo~A severely limited wide-area blind surveys in the region. For instance, NVSS, RACS low, TGSS, and VLSSr all have a local rms noise many times higher than the nominal one and a reduced fidelity due to the presence of many artefacts coming from an improper calibration and deconvolution of M87. Having an ad hoc strategy to subtract the effect of M87 during calibration, the ViCTORIA surveys can achieve higher levels of sensitivity and fidelity (see e.g. Fig.~\ref{fig:contcomp}). 

\section{Science cases and preliminary results}
\label{sec:science}

\subsection{Galaxy evolution}

The Virgo cluster is an ideal laboratory for studying the effects of the environment on galaxy evolution. Located only 16.5~Mpc away, it allows us to study all galaxy populations down to the dwarf regime with unprecedented sensitivity and at an unprecedented angular resolution. This is crucial for understanding the effects of the different perturbing mechanisms (gravitational and hydrodynamical) on the different gas components and on the star formation process down to the scale of individual giant molecular clouds and \Hii{} regions. Dwarf systems are of fundamental importance since they have a shallow gravitational potential well; they are thus the most easily perturbed objects in any kind of interaction.

\paragraph{HI gas:} The MeerKAT data at 21 cm provide us with a complete, \Hi-selected sample of galaxies in a rich cluster. These data are thus perfectly suited to reconstructing the \Hi{} mass function and comparing it with those derived in other environments such as clusters, groups, and the fields that recent \citep{Zwaan1997, Zwaan2005, Rosenberg2002, Martin2010, Moorman2014, Said2019}, ongoing, and future surveys \citep[e.g. WALLABY or the MeerKAT Fornax Survey;][]{Koribalski2020, Serra2023} will soon provide. The complete census of the \Hi{} content of gas-rich systems down to $M_{\Hi} \simeq 2 \times 10^6$~M$_{\odot}$ on a statistically significant sample (more than 300 objects) allows us to reconstruct the main \Hi{} scaling relations of galaxies down to the dwarf regime and compare them to those derived for field objects \citep[e.g.][]{Catinella2018, Saintonge2022} to quantify the impact of the environment on the atomic gas content of cluster systems \citep[e.g.][]{Cortese2011}.

\begin{figure}
        \centering 
        \includegraphics[width=.49\textwidth]{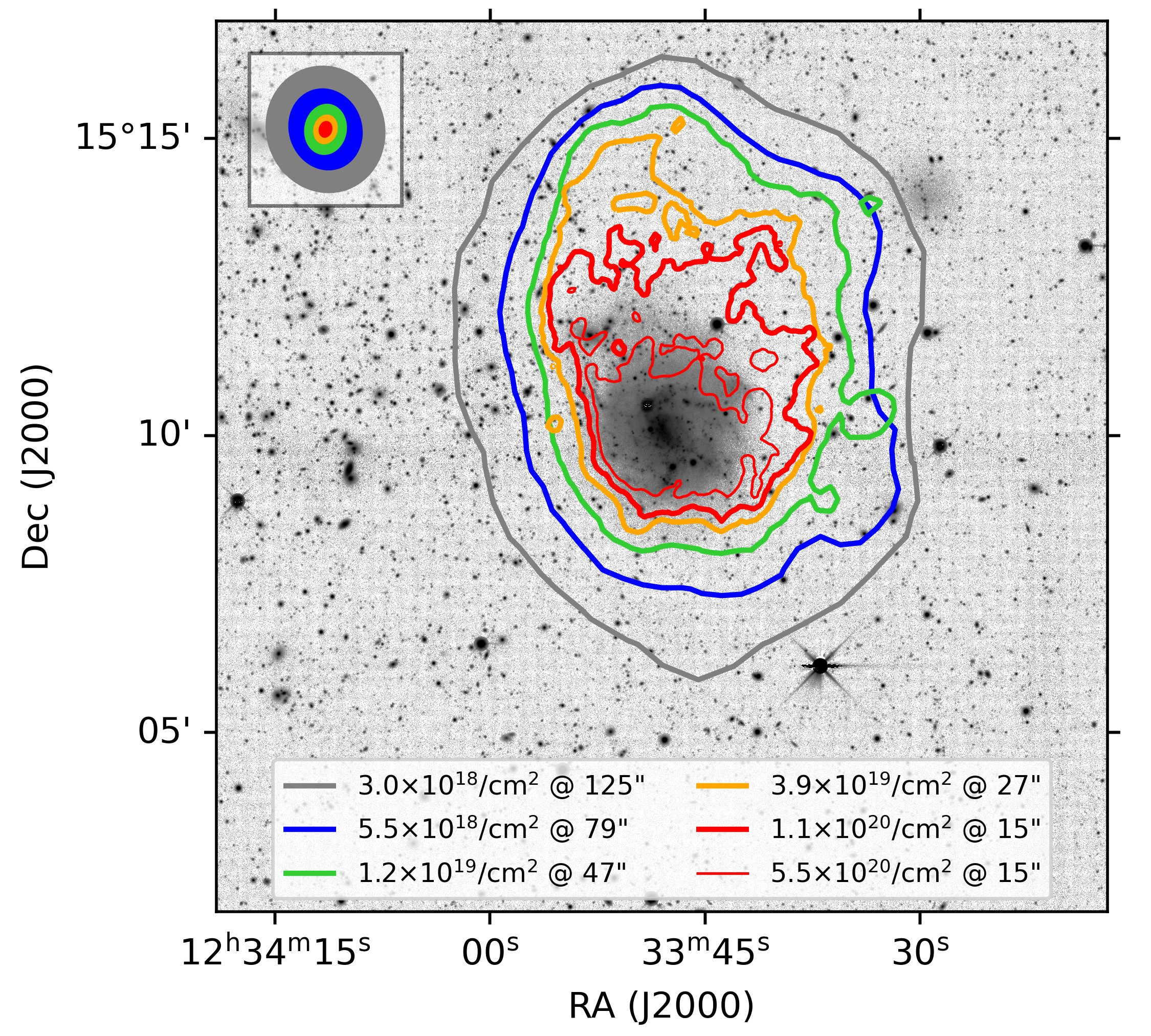}
        \caption{\Hi{} gas distribution of NGC~4523. For each resolution, we show the lowest reliable contour ($3\sigma$ over a line width of 25~\kms). The contours are overlaid on an $r$-band image downloaded from the Legacy Survey database. The contour levels and resolutions are listed in the legend at the bottom, and the resolutions are shown in the top left corner. At the highest resolution of $15\arcsec$, we also show an additional contour at a $5\times$ higher column density than the lowest reliable contour. These \Hi{} images reach column densities a factor of $\sim \sqrt{2}$ better than those reached during the pilot observations published in \citet{Boselli2023c}.}
        \label{fig:NGC4523}
\end{figure}

Distributed on a thin ($1-5$~kpc) and low-column-density $N(\Hi) \simeq 10^{19}-10^{20}$~cm$^{-2}$) disc, \Hi{} gas is a baryonic component that is less bound to the gravitational potential well of galaxies. For this reason, it is easily removed in any kind of interaction. The presence of tails of cold \Hi{} gas are thus ideal tracers of galaxies undergoing a perturbation \citep[e.g.][]{Chung2007}. They can also be used to identify the dominant perturbing mechanism, which is gravitational when associated with an asymmetry in the old stellar component and hydrodynamic when only on the gaseous component \citep{Boselli2022, Boselli2023a}. \Hi{} -resolved data are fundamental in the study of individual objects when combined with multi-frequency data and compared with tuned models and simulations \citep[e.g.][]{Vollmer2008, Vollmer2009b, Vollmer2012, Vollmer2018, Vollmer2021, Boselli2018b, Boselli2018c}. As an example, a pilot study using \Hi{} data gathered at the beginning of this project allowed us to identify a galaxy undergoing a perturbation (NGC~4523), recognise ram pressure stripping as the dominant mechanism, and study the effects of the perturbation on the kinematics of the gas in detail \citep[][]{Boselli2023b}. Using all data of the Virgo MeerKAT survey, we reach an $\sim \sqrt{2} \times$ better sensitivity level than in the pilot study, and Fig.~\ref{fig:NGC4523} shows the updated \Hi{} image of that galaxy. The data provided by the full survey of the cluster will enable the extension of this kind of analysis to several tens of objects and thus give us a more representative view of the complex interaction between galaxies and their surrounding environment.

Combined with data available for the other gas phases (cold molecular, ionised, hot), the \Hi{} data will be crucial for understanding the fate of the stripped gas within the hot ICM. Several observations and simulations suggests that the gas is stripped from the galaxy disc mainly in its cold atomic phase. Once mixed with the hot ICM, it warms up after energy transfer via shocks, heat conduction, magneto-hdyrdodynamic waves, becoming first ionised than hot gas \citep[e.g.][]{Tonnesen2010, Tonnesen2011, Tonnesen2012, Fossati2016, Boselli2016a, Boselli2021, Boselli2022}. Whenever the gas density in the tail is sufficiently high, the gas can collapse to form giant molecular clouds and then new stars. Part of the gas, the one located not far from the galaxy, can fall back on the disc once the galaxy reaches the cluster outskirts \citep{Vollmer2001}. The remaining gas is definitively stripped, and after mixing with the surrounding medium, it contributes to the pollution of the ICM \citep{Longobardi2020a, Longobardi2020b}.
The 21 cm observations will thus be crucial to identify tails of stripped gas as those already observed in seven galaxies in the VIVA survey \citep{Chung2007} or in NGC~4523 where the transformation of the stripped material can be finally studied with unprecedented statistics. The data can also be used to search for free floating \Hi{} clouds not associated to any stellar disc, generally called almost dark galaxies, as those detected by the ALFALFA survey \citep{Haynes2007, Kent2007, Koopmann2008, Kent2009, Minchin2007, Minchin2019}, which might be residual of past interactions (see Fig. \ref{fig:k09clouds}).

Finally, the kinematic information that the \Hi{} data cubes will provide for the extended tails of stripped gas can be used to reconstruct the 3D-orbital parameters of galaxies within the Virgo cluster, as done with NGC~4569 by Sarpa et al. (in prep.).

\paragraph{Continuum:} The sensitivity of the LOFAR and MeerKAT surveys will allow us to measure the radio-continuum luminosity function of the cluster at three different frequencies, significantly improving the one now available at 20~cm and based on the shallow NVSS data \citep{GB1999a}. As for the \Hi{} mass function, the radio continuum luminosity functions will be compared to those that all sky surveys will soon provide for galaxies in different density regions from the general field, voids, groups, and rich clusters of galaxies \citep[e.g.][]{Norris2021, Shimwell2022}. The same luminosity functions will be references for high-redshift studies. The radio-continuum data can be used to trace the mean statistical properties of galaxies in a rich nearby cluster (scaling relations) and compare them to those of similar galaxies in the field to quantify the impact of the environment on the radio emission down to the dwarf population. It will be possible to check whether the radio emission of cluster galaxies is increased with respect to those of field objects on global and local scales, as first noticed at 1400~MHz by \citet{Gavazzi1991} in the Coma cluster and later confirmed in the Virgo cluster at other frequencies \citep{Niklas1995, GB1999a, Murphy2009}. In particular, the different radio polarisation properties will be used to measure the ordered magnetic field over the disc of the detected galaxies and thus test whether the claimed compression of B is at the origin of the observed increased radio emission of cluster galaxies \citep{Gavazzi1991, VolkXu1994, GB1999a, GB1999b, Edler2024}.

The discovery of head-tailed radio galaxies in rich clusters was the first observational evidence of ram pressure exerted by the external medium on galaxies moving at high velocity within it \citep[e.g.][]{GunnGott1972}. Since then, extended radio-continuum tails have been observed in several cluster galaxies at different radio frequencies \citep[e.g.][]{Gavazzi1995, Roberts2021a, Roberts2021b, Roberts2022} and are becoming efficient tools to identify objects undergoing an external perturbation. Measuring the magnetic field in the stripped material \citep{Vollmer2021, Ignesti2022}, here made possible by polarisation studies, is important to understand the star formation process in this extreme environment. Indeed, several tuned magneto-hydrodynamic simulations suggest that in the presence of magnetic-field matter remains confined, meaning relatively high gas column densities are necessary for the collapse of gas into giant molecular clouds where stars are formed \citep{TonnesenStone2014, Ruszkowski2014}.

The multi-frequency radio-continuum data, which will be available for several hundreds objects, will be combined with those already gathered at other frequencies (H$\alpha$, UV, far-IR) to study the star formation process on different timescales \citep[e.g.][]{Boselli2015, Vollmer2022, Edler2024}. Typical timescales will also be inferred by combining the three ViCTORIA datasets and measuring the steepening of the radio spectral slope in the tails of stripped material \citep{Vollmer2004, Chen2020, Vollmer2021, Ignesti2023, Roberts2024}. Combined with other temporal tracers, such as the ageing of the stellar populations, these data will be crucial to measuring the typical timescale for gas stripping and quenching of the star formation process; these are other important parameters necessary for the identification of the dominant perturbing mechanism in rich environments \citep[e.g.][]{Boselli2022}.

\subsection{AGNs and M87}

\begin{figure*}[ht!]
        \centering 
        \includegraphics[width=.33\textwidth]{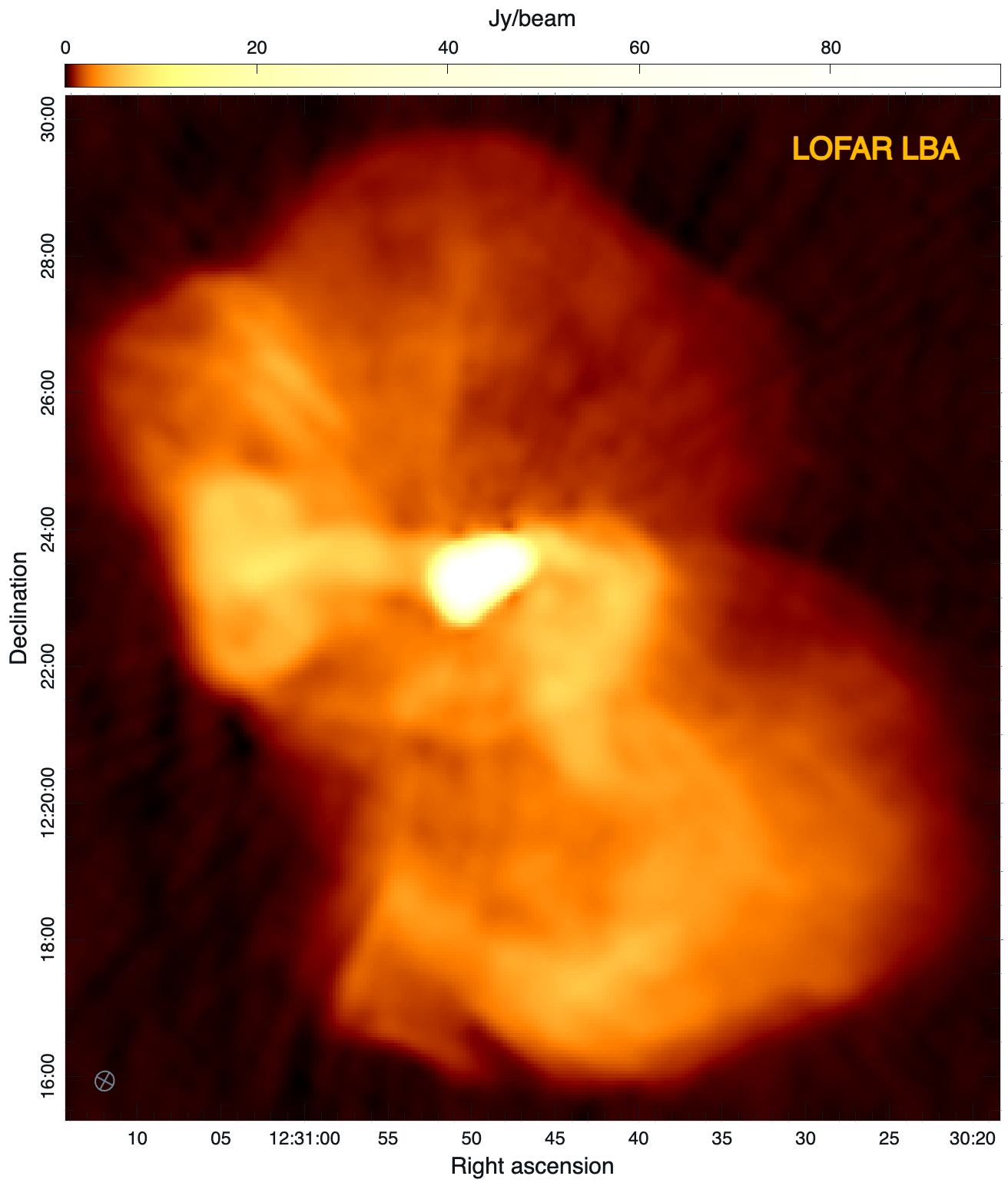}       \includegraphics[width=.33\textwidth]{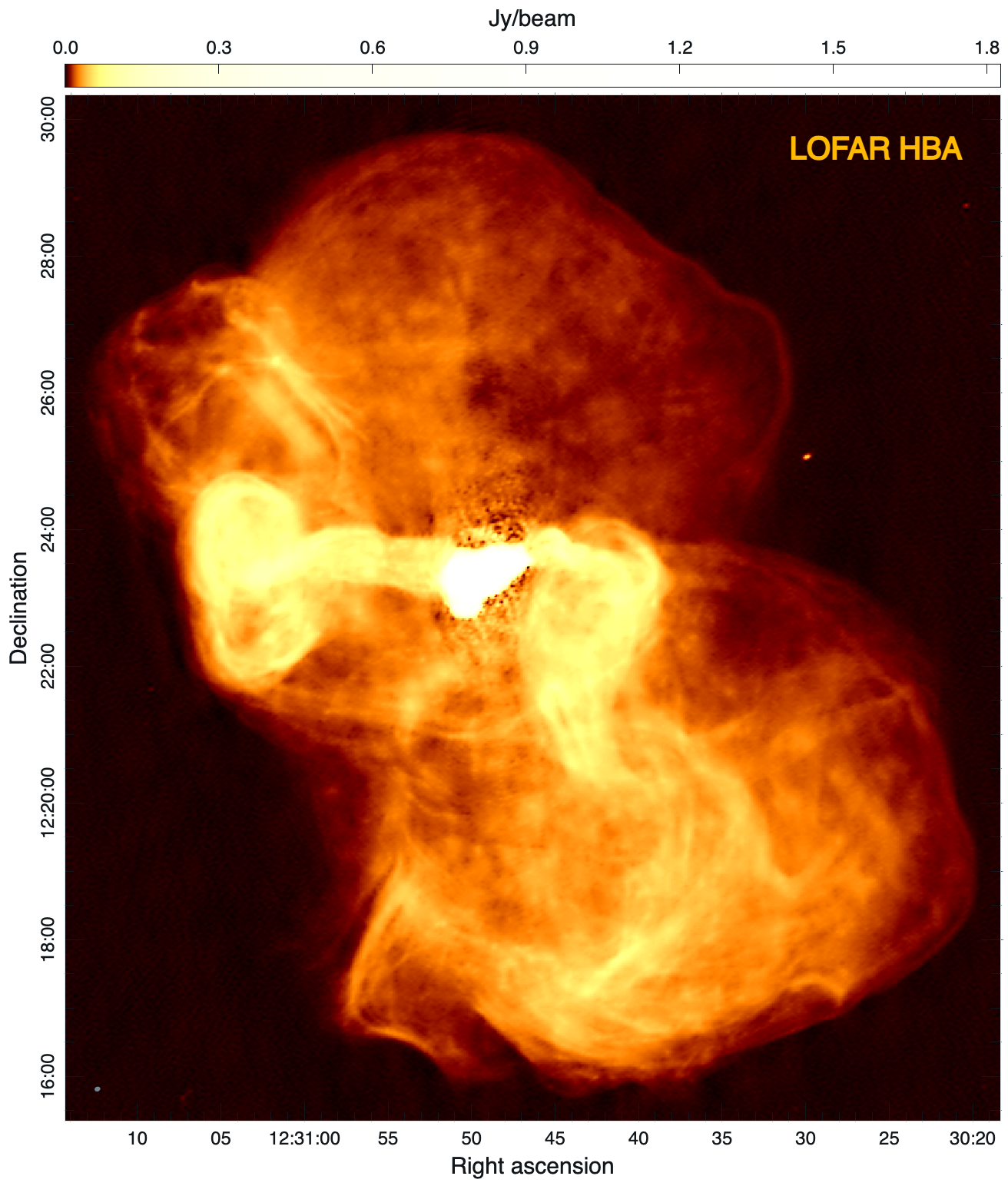}
        \includegraphics[width=.33\textwidth]{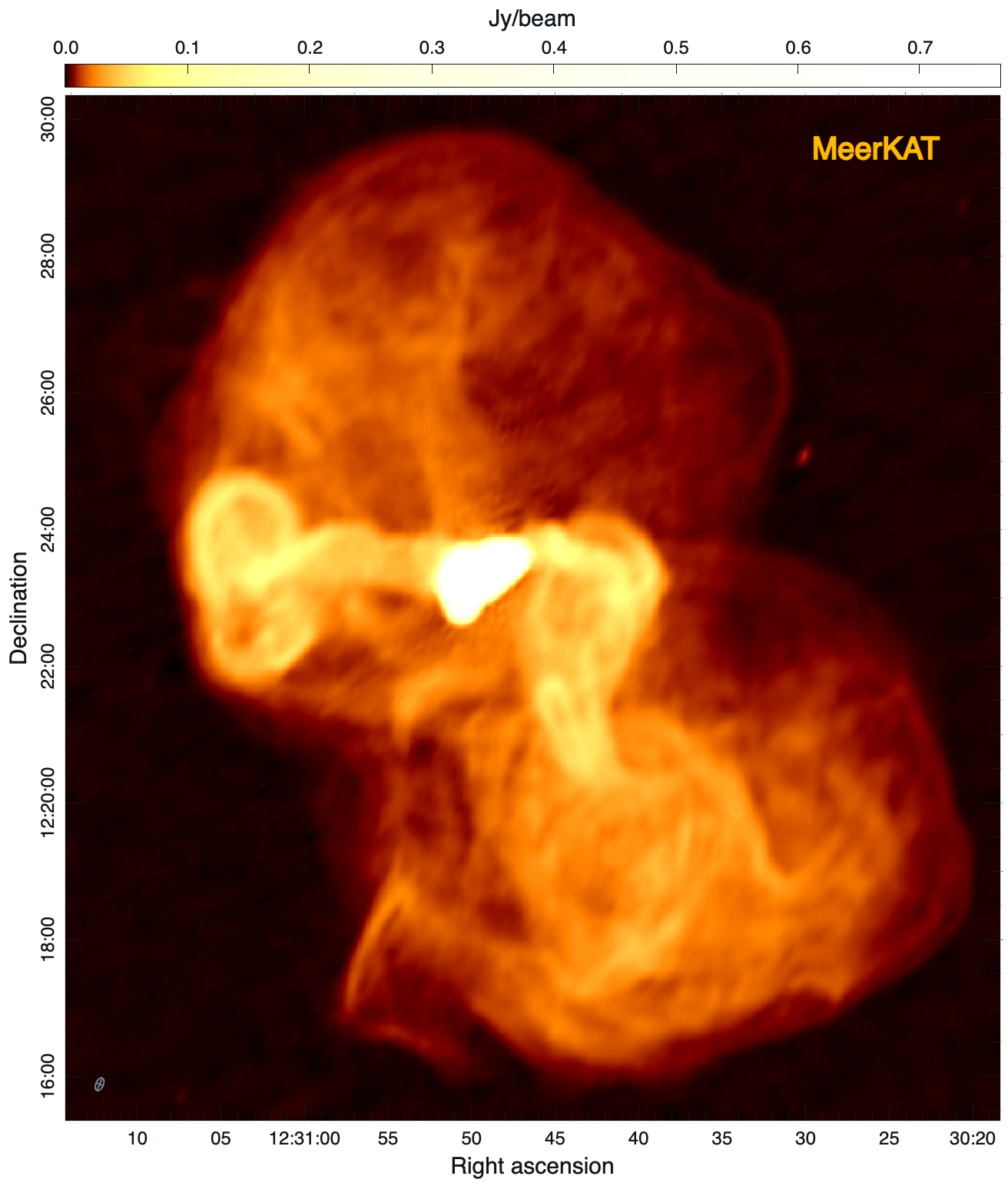}
        \caption{Virgo A (the radio emission associated with M87) as seen by the three ViCTORIA surveys (de Gasperin et al. in prep.). From left to right: LOFAR LBA (54 MHz, beam: \beam{16}{12}), LOFAR HBA (144 MHz, beam: \beam{4}{3} including LOFAR international stations), and MeerKAT L band (1284 MHz, beam: \beam{12}{6}). The beam size is shown in the bottom left corner of each image.}
        \label{fig:m87}
\end{figure*}

Usually, only a few AGNs are detected in each galaxy cluster. Thanks to the combination of observational depth and cluster distance, we will collect the first significant statistics of active nuclei in a galaxy cluster to compare their nuclear (instantaneous) and extended (integrated) radio emission with the galaxy properties and location across the cluster. AGNs will also be detected in background clusters and groups. Based on nuclear spectroscopy \citep{Cattorini2023} and X-ray observations \citep{Gallo2010, Hou2024}, the known number of AGNs in the survey footprint is 81. Initial results are presented in \citet{Spasic2024}.

\begin{figure}[ht!]
        \centering 
        \includegraphics[width=.5\textwidth]{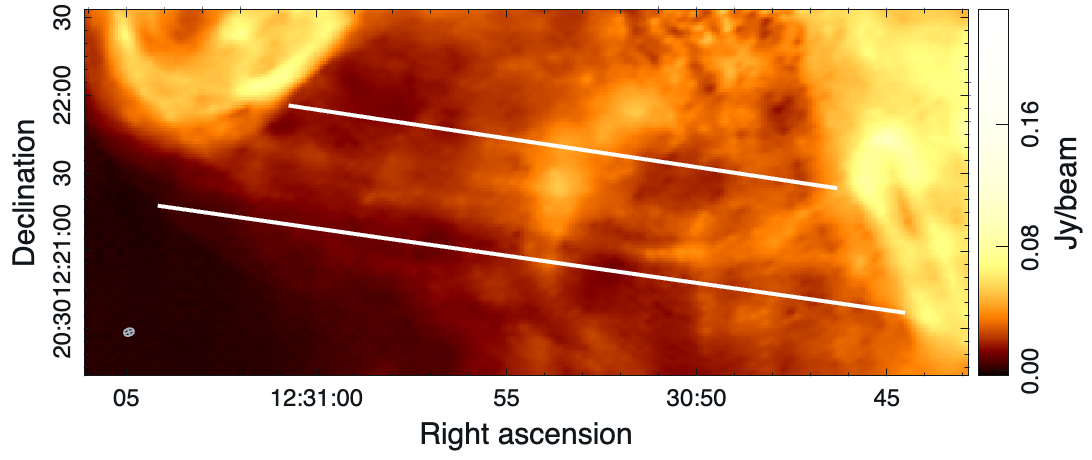}
        \caption{Extract from central panel of Fig.~\ref{fig:m87}, where the collimated synchrotron thread is visible and joins the two 'ears'.}
        \label{fig:m87cst}
\end{figure}

At the centre of the Virgo cluster resides the notorious M87 galaxy, whose radio emission is also called Virgo A. At the centre of M87 resides M87$^*$, the first supermassive black hole imaged with the Event Horizon Telescope \citep{EventHorizonTelescopeCollaboration2019}. Its jets have also been meticulously studied from the formation region \citep{Lu2023} up to their extension \citep{Owen1999}.
The source has also been extensively observed in the X-ray \citep{Forman2007, Million2010}, where \textit{Chandra} and \textit{XMM-Newton} revealed, along with hot gas at $2-3$~keV, the presence of extended gas filaments at lower temperature and entropy which are closely related to the radio-emitting structures. Two AGN-driven shocks ($M \simeq 1.2$) have been found at distances of about 3~kpc and 14~kpc, respectively, from the central AGN \citep{Forman2007, Million2010}. These repeated shocks might have distributed part of the energy necessary to offset the radiative cooling of the ICM. Finally, several bubbles of plasma and cavities excavated by injections of relativistic particles have been detected both in the inner region and in the extended halo \citep{Forman2007}. Bubbles and cavities contain even larger amounts of energy, which can also contribute to gas heating \citep{Churazov2001, Zhang2018}.

The AGN itself and signs of its interaction with the ICM have also been seen in the radio band \citep{Owen2000, deGasperin2012}. The morphology of this radio source cannot be easily classified as FRI/FRII, and it is characterised by a compact and bright inner region that contains the jet, an eastern lobe with a characteristic mushroom shape, and a south-western lobe that bends from the direction of the jet. The central region and the lobes are surrounded by two overlapping spherical halos of low brightness (about 40~kpc across). High-resolution maps and Faraday rotation data of the inner cocoon showed that the magnetic field (and therefore the radio emission) is highly filamentary \citep{Owen2000}. Interestingly, radio filaments twist and wrap around the coldest metal-rich X-ray emitting gas, suggesting a complex interplay between the radio (non-thermal) and the X-ray (thermal) plasmas \citep{Forman2007}.

Images obtained from the ViCTORIA project surveys (see Fig.~\ref{fig:m87}) provide the highest resolution images of the Virgo A extended structure (\beam{4}{3} at 144 MHz from LOFAR HBA data (including international stations), but also the images with the highest resolution at 54 MHz (beam: \beam{16}{12}) and 1284 MHz (beam: \beam{12}{6}) to date. The combination of these data will be exploited to study detailed spectral properties of the source (de Gasperin et al. in prep.). The data show evidence of complex filamentary structures filling up the volume of the lobes. Furthermore, the west plasma flow is now more similar to the famous ear to the east, suggesting that Rayleigh–Taylor instabilities are driving the shapes of both. A filamentary structure joining the two ears, likely due to a stretched or a reconnected magnetic field, also appear well visible below the central cocoon (see Fig.~\ref{fig:m87cst}). These elongated filaments (or collimated synchrotron threads; CST) connecting distant lobes of radio galaxies were detected in one other case and still have an unclear origin \citep{Ramatsoku2020}.

\subsection{Foreground science} 

The spectral domain chosen for the 21 cm observations covers a wide range of velocities, from the Milky Way up to $z \sim 0.6$. This is a unique dataset in terms of sensitivity and angular resolution used to study the \Hi{} gas distribution of the Galaxy at high galactic latitude ($b \geq 70\deg$) on a $\simeq 100$~deg$^2$ contiguous field. 
The \Hi{} data can be combined with those already available at other frequencies for this sky region to study the relation between the atomic gas distribution (GALFA, \citealt{Peek2011}; ALFALFA, \citealt{Bianchi2017}), the ionised gas distribution (WHAM, \citealt{Reynolds1998}; VESTIGE, \citealt{Boselli2018a}), and that of cold dust as traced by far-IR emission (IRAS, Planck, Herschel) or by the scattered light detected in the UV bands during the GUViCS survey \citep{Boissier2015}. The comparison done in  Sect. 3.1 clearly indicates that the sensitivity of the \Hi{} MeerKAT data is comparable to the one reached by ALFALFA at low angular resolution. It is thus likely that all the diffuse and extended features studied by \citep{Bianchi2017}
will be detected in our new set of data. The \Hi{} data can also be used to look for high-velocity clouds as those discovered during the GALFA-\Hi, \citep{Saul2012} and ALFALFA \citep{Adams2013, Bellazzini2015} surveys. As discussed in Sect. 3.1, the \Hi{} MeerKAT data have not only the sensitivity to detect these features, but also the angular resolution to study their nature in great detail.

Large-scale extended emission of galactic origin can be picked up more easily at lower frequencies, where the interferometers can detect larger patches of emission without resolving them out. Data from the LOFAR HBA Virgo Survey showed that we can detect large stripes of galactic emission crossing the survey footprint \citep{Edler2023}. Based on their location and orientation, the authors suggest an association with the North Polar Spur (Loop I), a Galactic spherical structure that can be either local ($d \sim 100$~pc) or originating from the Galactic centre.

\subsection{Background science}

Galaxies belonging to large scale structures will be detected during the ViCTORIA survey. The most famous of these structures, the Great Wall \citep{Geller1989}, falls within the footprint of the survey. The Great Wall has a filamentary structure typical of superclusters, extending over hundreds of degrees on the sky at a fairly common distance of $\sim 80-140$~Mpc. Similar structures are also present at further distances. At the sensitivity level of the survey (noise level of 0.6~\mjybeam{} per 5.5~\kms channel), the $5\sigma$ $M_{\Hi}$ detection limit assuming a \Hi{} line width of 200~\kms for a galaxy at a distance $D({\rm Mpc})$ is $M_\Hi = 2.3 \times 10^4$ M$_\odot \times D({\rm Mpc})^2$. For a typical distance of $D \simeq 100$~Mpc, we can detect galaxies down to $M_\Hi = 2.3 \times 10^8$~M$_\odot$. Detections of galaxies in the Great Wall in the background of Virgo have been already reported by \citet{Hoffman1995} and more recently by \citet{Haynes2018}. These galaxies will provide a comparison sample of fairly isolated systems to which the Virgo data will be compared with the purpose of understanding the effects of the cluster environment on galaxy evolution.

The sky in the direction of the Virgo cluster region has been covered with deep blind surveys at almost all frequencies (see Sect.~\ref{sec:complementary}). This region is thus an ideal laboratory in which to study galaxy evolution as a function of cosmic time with a multi-frequency statistical analysis. While not as deep as other well studied fields \cite[e.g. COSMOS,][]{Schinnerer2007, Smolcic2017}, this region is sufficiently wide ($\sim 100$~deg$^2$) to be free from cosmic variance, and thus it is ideally suited for studying the statistical distribution of galaxies (correlation function) and of their properties as a function of local density in different bands thanks to the NGVS photometric redshift catalogue of $\sim 3 \times 10^7$ sources \citep{Raichoor2014}. 


\subsection{Cluster magnetic field}

Following \citet{Rudnick2014}, and considering a $6\sigma$ detection threshold, we expect to detect 3677 polarised sources\footnote{MIGHTEE early science fields suggests that the number of polarised sources in the area might be underestimated by a factor 2 \citep{Taylor2024}.} from the MeerKAT Virgo Cluster Survey within the cluster region. About 43 should be detectable with the LOFAR HBA Virgo cluster Survey \citep{OSullivan2023}. Compared to the L band, RM values extracted from LOFAR HBA data have a higher level of precision. We do not expect to detect a significant number of polarised sources in the LBA survey due to depolarisation. Having 3677 RM measurements will enable the derivation of the magnetic field strength and structure by modelling the RM profile in the cluster \citep{Murgia2004, Bonafede2010} or the radial depolarization trend \citep{Osinga2022}. This will be made possible thanks to the complementary X-ray observations, which will be used to derive the thermal electron density profile \citep{McCall2024}. An accurate measure of the strength of the magnetic field is important to understand the origin of the large-scale radio emission and the relevance of this non-thermal component with respect to the thermal component.

\section{Conclusions}
\label{sec:conclusions}

In this paper, we outline the Virgo Cluster multi-Telescope Observations on Radio of Interacting galaxies and AGNs (ViCTORIA) Project and described its three constituents: 1. The LOFAR LBA Virgo Cluster Survey ($42-66$ MHz); 2. The LOFAR HBA Virgo Cluster Survey ($120-168$ MHz); and 3. The MeerKAT Virgo Cluster survey ($856-1712$ MHz). The data of all three surveys cover out to about $r_{200}$, including the main sub-clusters. The project will deliver three blind surveys of the Virgo cluster spanning the $42-1712$ MHz frequency range, polarisation data in the L band with rotation measure down to 120 MHz, a blind \Hi{} survey of the entire region, and a spectral index map of any emission detected in multiple bands.

The project showed its potential with some early results including a detailed study of the complex \Hi{} structure in NGC~4523 \citep{Boselli2023a}, the analysis of a sample of 17 ram-pressure-stripped galaxies in the Virgo cluster \citep{Edler2024}, and the study of the AGN emission from a sample of 12 early-type galaxies that included the discovery of large radio-filled cavities in the famous M49 galaxy \citep{Spasic2024}. In this paper, we show some forthcoming results, presenting the highest resolution maps ever obtained of the radio emission surrounding M87 in the survey's three frequency ranges. We show that the whole lobes are filled with resolved filamentary structures, including the second known case of CST linking the two rising lobes.



\begin{acknowledgements}

FdG acknowledge the support of the ERC Consolidator Grant ULU 101086378.

The Low Frequency Array, designed and constructed by ASTRON, has facilities in several countries, that are owned by various parties (each with their own funding sources), and that are collectively operated by the International LOFAR Telescope (ILT) foundation under a joint scientific policy.

The MeerKAT telescope is operated by the South African Radio Astronomy Observatory, which is a facility of the National Research Foundation, an agency of the Department of Science and Innovation.

This research has made use of NASA's Astrophysics Data System.

This research has made use of SAOImage DS9, developed by Smithsonian Astrophysical Observatory.

\end{acknowledgements}


\bibliographystyle{aa}
\bibliography{library}


\end{document}